%% file: pub649.tex
\RequirePackage{xspace}
\documentclass[aps,prl,tightenlines,superscriptaddress,twocolumn]{revtex4-2}
\usepackage{xcolor}
\usepackage{graphicx}
\usepackage{dcolumn}
\usepackage{bm}
\usepackage{multirow}
\usepackage{amsmath}

\usepackage{orcidlink}
\usepackage{xurl}
\hypersetup{breaklinks=true}

\def \asy#1#2{\ensuremath{{}^{+#1}_{-#2}}\xspace}

\begin{document}
\title{Study of $\boldsymbol{\chi_{bJ}(2P)\rightarrow\omega\Upsilon(1S)}$ at Belle}
\input{pub649-orcid}

\begin{abstract}
We report a study of the hadronic transitions $\chi_{bJ}(2P)\to\omega\Upsilon(1S)$, with $\omega\to\pi^{+}\pi^{-}\pi^{0}$, using $28.2\times10^6~\Upsilon(3S)$ mesons recorded by the Belle detector. We present the first evidence for the near--threshold transition $\chi_{b0}(2P)\to\omega\Upsilon(1S)$, the analog of the near-threshold charm sector decay $\chi_{c1}(3872)\to\omega J/\psi$, with a branching fraction of $\mathcal{B}\big(\chi_{b0}(2P)\to\omega\Upsilon(1S)\big) = \big(0.55\pm0.19\pm0.07\big)\%$. We also obtain branching fractions of $\mathcal{B}\big(\chi_{b1}(2P)\to\omega\Upsilon(1S)\big) = \big(2.39\asy{0.20}{0.19}\pm0.24\big)\%$ and $\mathcal{B}\big(\chi_{b2}(2P)\to\omega\Upsilon(1S)\big) = \big(0.47\asy{0.13}{0.12}\pm0.06\big)\%$, confirming the measurement of the $\omega$ transitions of the $J=1,2~P$--wave states. The ratio for the $J=2$ to $J=1$ transitions is also measured and found to differ by 3.3 standard deviations from the expected value in the QCD multipole expansion.
\end{abstract}

\maketitle
\tighten

% \section{I. Introduction} \label{sec:introduction}
Recently, the hadronic transitions among heavy quarkonium $(Q\bar{Q}$, where $Q=c,b)$ states have been the focus of detailed study~\cite{Chen2008,Aubert2008,delAmoSanchez2010,Lees2011,Adachi2012,Tamponi2015,EGuido2017,Guido2018,Oskin2020,Ablikim2019,LHCb3872omega}. Of such transitions, those occurring near kinematic thresholds for the decay constitute a unique laboratory in which one can study the emission and hadronization of low-momentum gluons~\cite{Brambilla2011}. The observation of the near-threshold transition $\chi_{c1}(3872)\to\omega J/\psi$ by Belle~\cite{SKChoi2005}, subsequently confirmed by BaBar~\cite{delAmoSanchez2010}, BESIII~\cite{Ablikim2019}, and LHCb~\cite{LHCb3872omega}, is of particular interest. Although it is a narrow state ($\Gamma_{\chi_{c1}(3872)} = 1.19\pm0.21$~MeV~\cite{workman2022}) that lies nearly $8~\text{MeV}$ below the nominal kinematic threshold for production of an $\omega$ and $J/\psi$ meson, the observed branching fraction of $(5.0\pm1.9)\%$~\cite{workman2022} is nearly as large as that of the discovery channel $\chi_{c1}(3872)\to\pi^+\pi^- J/\psi$~\cite{Choi2003}. 

In the bottomonium $(b\bar{b})$ sector, the analogous $\omega\Upsilon(1S)$ final-state threshold lies between the $J=1$ and $J=0$ states of the $\chi_{bJ}(2P)$ triplet. With a nominal $Q$ value, calculated using the nominal masses \cite{workman2022}, of $Q = M_{\chi_{b0}(2P)} - M_{\Upsilon(1S)} - M_{\omega} = -10.5~\text{MeV}$, such a transition is expected to be suppressed by the substantially limited phase space. Although no experimental measurement of the $\chi_{b0}(2P)$ total width has been made, the current estimate from theory is 2.6~MeV \cite{Godfrey2015}. The broad nature of the $J=0$ state, coupled with the wide nature of the $\omega$ meson $(\Gamma_\omega = 8.68~\text{MeV~ \cite{workman2022}})$, provides a narrow phase space through which the $J=0$ transition may proceed through the low-energy tail of the $\omega$ line shape.

% In the bottomonium $(b\bar{b})$ sector, the analogous $\omega\Upsilon(1S)$ final--state threshold lies between the $J=1$ and $J=0$ states of the $\chi_{bJ}(2P)$ triplet, with the latter lying about $10.5~\text{MeV}$ below the nominal threshold.

In 2004, CLEO reported the first observation of the transitions $\chi_{bJ}(2P)\to\omega\Upsilon(1S)$ produced in radiative decays of $(5.81\pm0.12)\times10^6~\Upsilon(3S)$ mesons. The branching fractions of the $J=1$ and $J=2$ states were measured to be $(1.63\asy{0.35}{0.31}\asy{0.16}{0.15})\%$ and $(1.10\asy{0.32}{0.28}\asy{0.11}{0.10})\%,$ respectively~\cite{Pedlar2004}.  To date, no confirmation of the CLEO measurement has been made. Although no indication of a $J=0$ signal was seen by CLEO, Monte Carlo (MC) simulation of $\chi_{b0}(2P)$ transitions to an $S$--wave $\omega\Upsilon(1S)$ indicates that the decay may be observed, though in such transitions the $\omega$ lineshape is distorted due to the presence of the nearby kinematic threshold. In the CLEO analysis, a kinematic fit of the $\pi^{+}\pi^{-}\pi^{0}$ final state was constrained to the nominal $\omega$ mass with a cut on the resultant $\chi^2$. It is possible that $J=0$ signal events were suppressed as a result, due to the anticipated distortion in the $\omega$ mass lineshape. In the charmonium $(c\bar{c})$ sector, BaBar has reported a similar distortion in the lineshape of the $\omega$ mass for $\chi_{c1}(3872)\to\omega J/\psi$ events~\cite{delAmoSanchez2010}.

In this Letter, we report the first evidence of the near--threshold transition $\chi_{b0}(2P)\to\omega\Upsilon(1S)$ with significance in excess of $3.2$ standard deviations, and confirm CLEO's measurement~\cite{Pedlar2004} of the corresponding transitions of the $J=1$ and $J=2$ states. The $\chi_{bJ}(2P)$ are produced in radiative transitions of the $\Upsilon(3S)$ at the Belle experiment. The $\Upsilon(1S)$ is reconstructed in decays to a pair of high-momentum leptons $(e,\mu)$, while the $\omega$ is reconstructed in its decay to $\pi^+\pi^-\pi^0$, with $\pi^0\to\gamma\gamma$. As a normalization sample, we reconstruct the decay $\Upsilon(3S)\to\pi^{+}\pi^{-}\Upsilon(1S)$ in the final state $\pi^{+}\pi^{-}\ell^{+}\ell^{-}$, which has a branching fraction of $(2.1\pm0.3)\times10^{-3}$~\cite{workman2022}.

The cascade branching ratios,
\begin{equation}
    \label{eqn:rJ1}
    r_{J/1} = \frac{\mathcal{B}\big( \Upsilon(3S) \to\gamma\chi_{bJ}(2P) \to\gamma\omega\Upsilon(1S) \big)}{\mathcal{B}\big( \Upsilon(3S)\to\gamma\chi_{b1}(2P)\to\gamma\omega\Upsilon(1S) \big)},
\end{equation}
are also measured and compared with the expectation from the multipole expansion model of quantum chromodynamics (QCDME)~\cite{MVoloshin2003}, which we calculate using the current world averages of bottomonium state masses (or mass differences, where more precisely measured) and branching fractions~\cite{workman2022}.

% \section{II. Data Samples and Detector} \label{sec:data_samples_and_detector}
We analyze data samples corresponding to an integrated luminosity of $2.9~\text{fb}^{-\text{1}}$ and $513.0~\text{fb}^{-\text{1}}$ recorded near the $\Upsilon(3S)$ and $\Upsilon(4S)$ resonances, respectively, by the Belle detector~\cite{Abashian2002} at the KEKB asymmetric--energy $e^+e^-$ collider~\cite{Kurokawa2003,KEKB}. We also study a sample, referred to as the off--resonance sample, collected about $60~\text{MeV}$ below the $\Upsilon(4S)$ resonance, totalling $56.0~\text{fb}^{-\text{1}}$. The number of $\Upsilon(3S)$ in the combined dataset is determined from a reconstruction of $\Upsilon(3S)\to\pi^+\pi^-\Upsilon(1S)[\ell^+\ell^-]$ to be $(28.2\pm0.9)\times10^6$. Decays of $\Upsilon(3S)$ mesons in data accumulated above the $\Upsilon(3S)$ resonance are assumed to come from initial state radiation (ISR) by the $e^+e^-$ pair. No attempt has been made to reconstruct the ISR photons, which typically emerge at small angles to the beampipe~\cite{benayoun1999}.

The Belle detector, described in detail elsewhere~\cite{Abashian2002,Brodzicka2012}, is a large-solid-angle magnetic
spectrometer that consists of a silicon vertex detector (SVD),
a 50-layer central drift chamber (CDC), an array of
aerogel threshold Cherenkov counters (ACC), 
a barrel-like arrangement of time-of-flight
scintillation counters, and an electromagnetic calorimeter
(ECL) composed of CsI(Tl) crystals located inside 
a superconducting solenoid coil that provides a 1.5~T
magnetic field. The ECL is divided into three regions spanning the polar angle $\theta$. The $z$ axis is taken to be the direction opposite the $e^{+}$ beam. The ECL backward endcap, barrel, and forward endcap, cover ranges $\cos\theta\in[-0.91,-0.65]$, $[-0.63,0.85]$, and $[0.85,0.98]$, respectively. An iron flux-return located outside of the coil is instrumented with resistive plate counters to detect $K_L^0$ mesons and muons (KLM). The data collected for this analysis used an inner detector system with a $1.5~\text{cm}$ beampipe, 
a 4-layer SVD, and a small-inner-cell CDC.

A set of event selection criteria are devised to optimize the retention of signal events while suppressing backgrounds from misreconstructed $\pi^{0}$ decays, resonant $b\bar{b}$ decays, and nonresonant (continuum) production of other quark and lepton species. For all optimizations, we employ the figure of merit $S/\sqrt{S+B}$, where $S$ and $B$ denote the number of signal and background events, respectively. To study these criteria and their associated reconstruction efficiencies, MC simulated events are produced. MC samples containing an admixture of signal $\chi_{bJ}(2P)\to\omega\Upsilon(1S)$ and other $b\bar{b}$ events are prepared assuming the CLEO-measured branching fractions \cite{Pedlar2004,workman2022}. As no measurement of the $J=0$ signal transition exists, it is conservatively taken to be small and equivalent to the $J=2$ branching fraction. MC events are generated with the E{\scriptsize VT}G{\scriptsize EN}~\cite{Lange2001} package. Radiative transitions among $b\bar{b}$ states are assumed to be E1 (i.e.\ electric dipole) transitions and have been generated according to their helicity amplitudes~\cite{Jacob1959}. Di--pion transitions among the $\Upsilon(3S)$, $\Upsilon(2S)$, and $\Upsilon(1S)$ states are modeled according to the matrix elements reported in Ref.~\cite{Hennessy2007}. All other hadronic transitions among $b\bar{b}$ states are modeled with phase space. Final state radiation effects are modeled by {\sc P{\scriptsize HOTOS}}~\cite{EBarberio94}. The Belle detector response is simulated with GEANT3~\cite{GEANTBrun}. The time-varying detector performance and accelerator conditions are taken into account in the simulation.

% \section{III. Event Selection} \label{sec:event_selection}
Reconstructed charged particles are required to originate within $2.0~\text{cm}$ of the interaction point (IP) along the $z$ axis and within $0.5~\text{cm}$ in the transverse plane. Tracks whose momentum exceeds $4~\text{GeV}$\cite{unitsFootnote} measured in the center--of--mass (c.m.) frame are preliminarily identified as leptons, and pairs of such tracks are combined to form $\Upsilon(1S)$ candidates if their invariant mass lies within the range $M(\ell\ell)\in[9.0,9.8]~\text{GeV}$.

A likelihood $\mathcal{L}_i~(i=\mu,\pi,K)$ is ascribed to each track based on its signature in the KLM and its agreement with an extrapolation of the track from the CDC. The muon identification likelihood ratio is defined as $\mathcal{R}_\mu = \mathcal{L}_\mu / (\mathcal{L}_\mu + \mathcal{L}_\pi + \mathcal{L}_K)$ \cite{MuonIDAbashian2002}. A similar electron identification likelihood ratio is defined as $\mathcal{R}_{e} = \prod\limits_{i=1}^{n}\mathcal{L}_{e}^{i} / \left( \prod\limits_{i=1}^{n}\mathcal{L}_{e}^{i} + \prod\limits_{i=1}^{n}\mathcal{L}_{\bar{e}}^{i}\right)$, where $\mathcal{L}_e^i~(\mathcal{L}_{\bar{e}}^i)$ are the likelihoods associated with $i$-th measurements from the CDC, ECL, and ACC, with the assumption that the track is (not) an electron~\cite{Hanagaki2002}. Both lepton candidates are identified as muons if $\mathcal{R}_\mu > \mathcal{R}_e$ for either of them; otherwise, they are considered as electrons.

QED continuum processes of the form $e^+e^-\to q\bar{q}$, where $q=u,d,s,c$, may mimic our signal. Due to the relatively small production cross section of our signal in the $\Upsilon(4S)$ dataset, the signal purity is far lower than that of the on--resonance $\Upsilon(3S)$ dataset. To suppress the higher backgrounds present in $\Upsilon(4S)$ data, the flavor of $\Upsilon(1S)\to\ell^{+}\ell^{-}$ is positively reconstructed by requiring that both leptons satisfy $\mathcal{R}_e>0.2$ or $\mathcal{R}_\mu>0.2$. The identification efficiency for a muon~(electron) pair passing the likelihood ratio criterion is approximately $92\%$. Moreover, the lepton momenta must satisfy $p_\text{c.m.}<5.25~\text{GeV}$, to avoid a peak from low multiplicity QED processes near $M(\Upsilon(4S))/2\simeq5.29~\text{GeV}$.

Due to the limited phase space to produce a pair of charged kaons in conjunction with a neutral pion in events with an $\Upsilon(1S)$ candidate, all low-momentum tracks in the c.m. frame, satisfying $p_\text{c.m.}<0.43~\text{GeV}$, are treated as pion candidates. Contamination from photon conversion to an $e^+e^-$ pair in detector material are suppressed by requiring that the cosine of the opening angle between oppositely charged pion candidates be less than 0.95. To reject events with misreconstructed tracks, events with multiple pairs of oppositely charged pions are rejected.

Photons are reconstructed from isolated clusters in the ECL that are not matched with a charged track projected from the CDC. To reject hadronic showers, the ratio of energy deposited in a $3\times3$ and $5\times5$ array of crystals centered on the most energetic one is required to exceed 90\%. Clusters with an electromagnetic shower width exceeding $6.0~\text{cm}$ are also rejected. To suppress beam--related backgrounds, photons are required to have an energy greater than 50, 100, and $150~\text{MeV}$ in the barrel, backward and forward endcaps, respectively.

Neutral pion candidates are formed from pairs of photons that satisfy $M(\gamma\gamma)\in[0.11,0.15]~\text{GeV}$, which is approximately 90\% efficient. To reject combinatorial background from misreconstructed $\pi^0$ candidates, we require that $p_\text{c.m.}^{\pi^{0}}\in[0.08,0.43]~\text{GeV}$. A kinematic fit is performed to constrain the invariant mass of each candidate to the current world average $\pi^0$ mass~\cite{workman2022}, and the best--candidate $\pi^0$ is selected according the smallest mass--constrained fit $\chi^2$. Studies in simulation indicate that the best--candidate selection rejects 45\% of the background from misreconstructed $\pi^0$ at a cost of 14\% of the signal. The $\omega$ candidate is reconstructed as the combination of the $\pi^0$ and the $\pi^+\pi^-$ pair; those satisfying the reconstructed $\omega$ 
 mass $M_{\omega}\in[0.71,0.83]~\text{GeV}$ are retained for further analysis.

Backgrounds from resonant di-pion bottomonium transitions may mimic our final state $\pi^+\pi^-\pi^0\ell^+\ell^-$. The largest source of contamination arises from decay chains containing $\Upsilon(2S)\to\pi^+\pi^-\Upsilon(1S)$, which may be produced through feed--down transitions $\big(\pi^+\pi^-,$ $\pi^0\pi^0$, or $\gamma\gamma$ via $\chi_{bJ}(2P) \big)$ of the $\Upsilon(3S)$ or directly via ISR. In the $\Upsilon(4S)$ dataset, additional contamination arises from transitions of the form $\Upsilon(4S)\to\pi^{+}\pi^{-}\Upsilon(2S)$, where the $\Upsilon(2S)$ decays to the $\Upsilon(1S)$ through similar feed--down transitions. To suppress such events, we develop a veto using the shifted mass difference 
\begin{equation}
    \label{eqn:dmpp}
    \Delta M_{\pi\pi} = M(\pi^{+}\pi^{-}\ell^{+}\ell^{-}) - M(\ell^{+}\ell^{-}) + m\big(\Upsilon(1S)\big),
\end{equation}
where the broad resolution of the di-lepton invariant mass is removed by subtracting the reconstructed mass of the leptons and adding back the current world average $\Upsilon(1S)$ mass~\cite{workman2022}. Di-pion transitions between $b\bar{b}$ states give rise to narrow peaks in the $\Delta M_{\pi\pi}$ distribution with a resolution of approximately $2~\text{MeV}$.

Background from $\Upsilon(3S)\to\pi^{+}\pi^{-}\Upsilon(2S)$ and $\Upsilon(3S)\to\pi^{+}\pi^{-}\Upsilon(1S)$ events are rejected at no cost in signal efficiency by requiring $\Delta M_{\pi\pi} > 9.83~\text{GeV}$ and $\Delta M_{\pi\pi} < 10.13~\text{GeV}$, respectively. A single veto is developed to suppress contamination from $\Upsilon(2S)\to\pi^{+}\pi^{-}\Upsilon(1S)$ and $\Upsilon(4S)\to\pi^{+}\pi^{-}\Upsilon(2S)$, which have nearly overlapping signatures in $\Delta M_{\pi\pi}$ as a result of the similar mass splittings between the $n=2,4$ and $n=1,2$ $\Upsilon(nS)$ states. A figure of merit optimization yields $\Delta M_{\pi\pi}\notin(10.017,10.290)~\text{GeV}$ for the $\Upsilon(3S)$ and off--resonance datasets and $\Delta M_{\pi\pi}\notin(10.014,10.030)~\text{GeV}$ for the $\Upsilon(4S)$ dataset. These requirements reject 92\% of resonant $b\bar{b}$ events at a cost not exceeding 12\% of the signal.  Table~\ref{tab:yields} summarizes the selection efficiency in each channel.

\begin{center}
    \begin{figure}[htb]
        \includegraphics[width=0.5\textwidth]{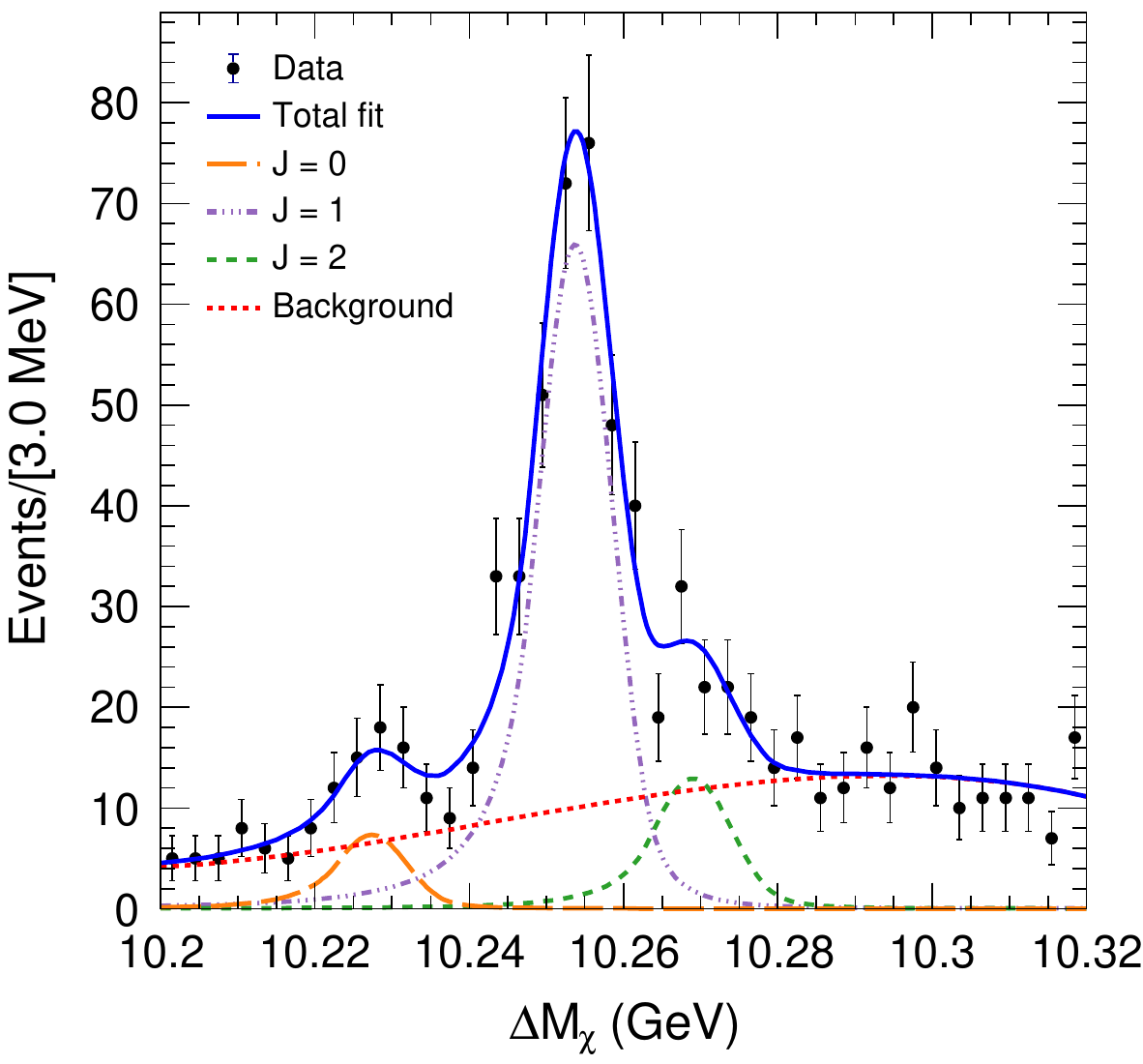}%
        \put(-112.8,131){\includegraphics[width=0.182\linewidth]{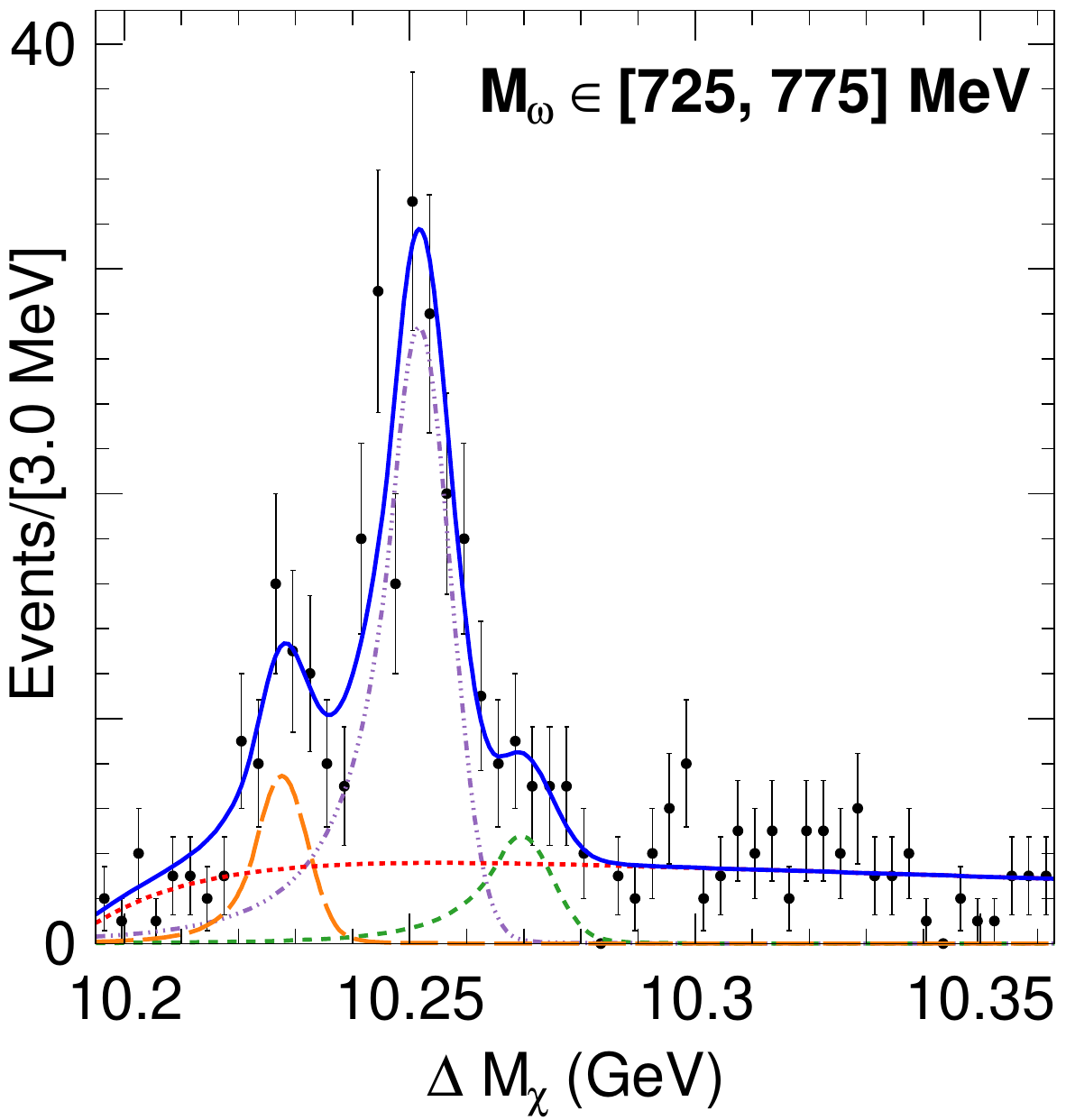}}
        \includegraphics[width=0.5\textwidth]{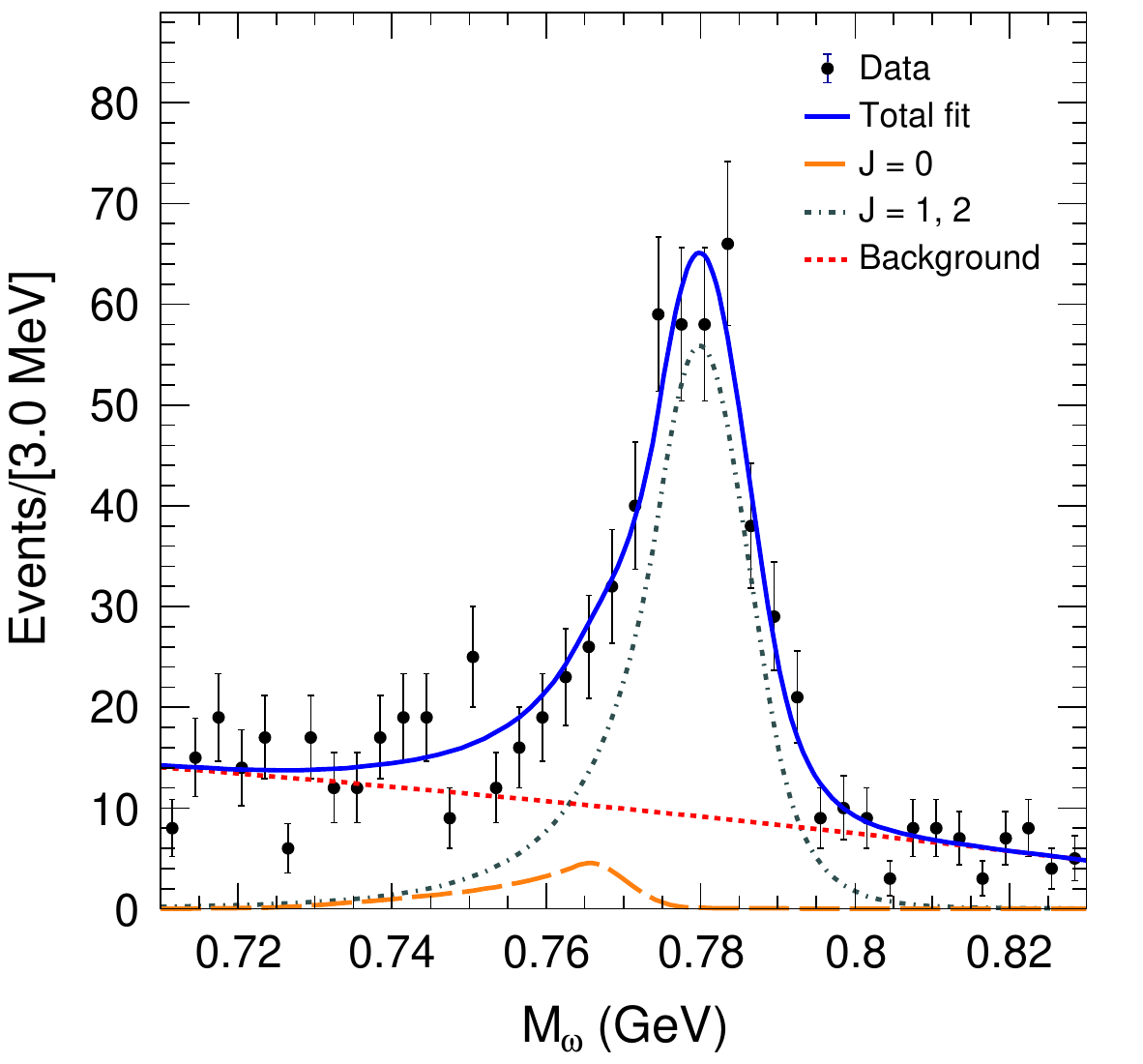}%
        \put(-212.8,131){\includegraphics[width=0.182\linewidth]{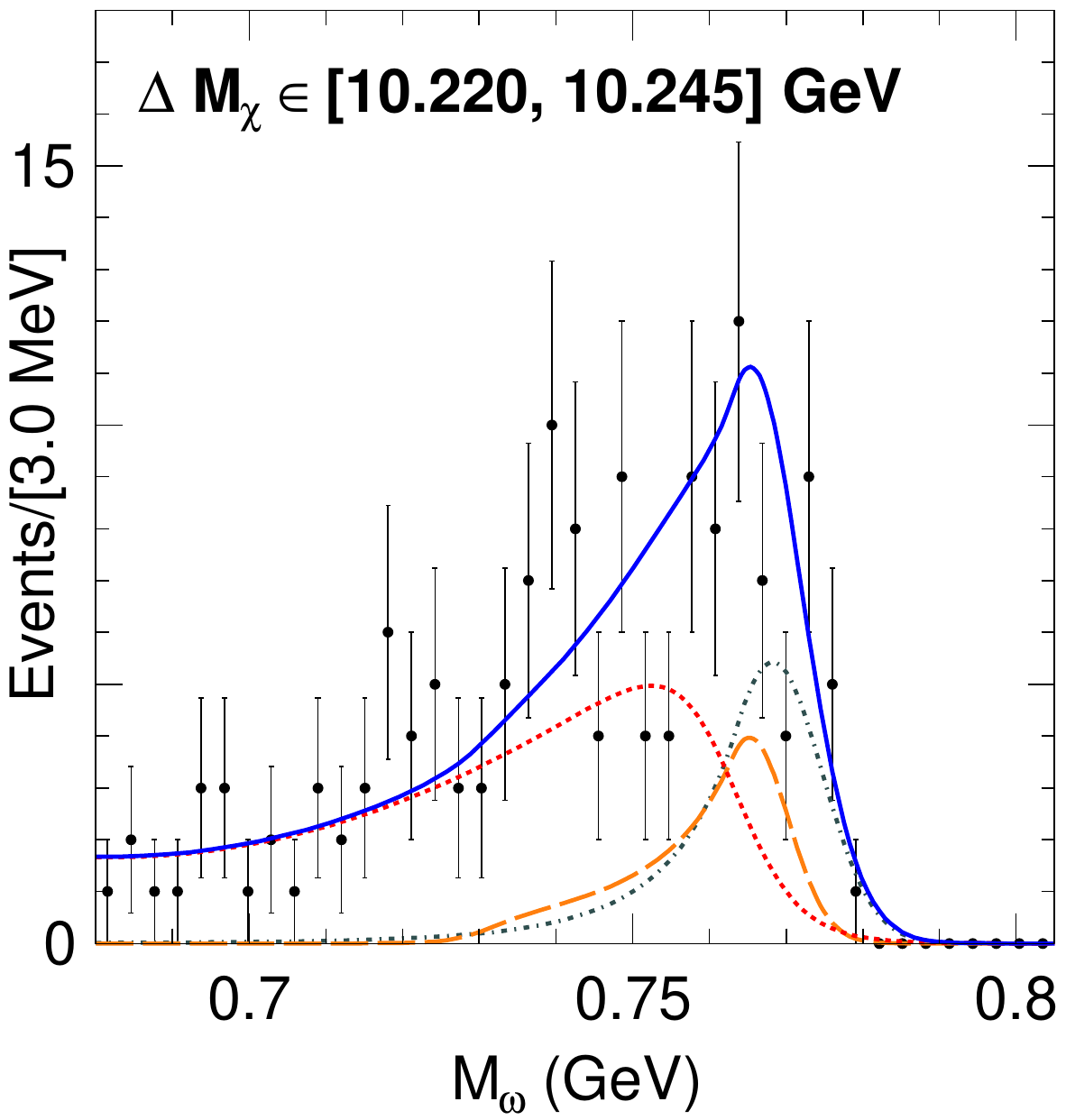}}

        \caption{Simultaneous fit to the $\Delta M_{\chi}$ (left) and $M_{\omega}$ (right) distributions for $\chi_{bJ}(2P) \to\omega\Upsilon(1S)$ candidates reconstructed in data. The inset panel on the left shows a fit to $\Delta M_{\chi}$ with only $\omega$ candidates that satisfy $M_\omega\in[725,775]$~MeV; the number of $J=0$ candidates extracted from the fit is consistent with the simultaneous fit result. The inset panel on the right shows a fit to the $M_{\omega}$ distribution for $\chi_{bJ}$ in the $J=0$ signal region, corresponding to $\Delta M_\chi \in [10.22, 10.245]~\text{GeV}$; note the similarity of the distorted $J=1,2$ lineshape to that of the $J=0$. The solid blue curves show the total fit and the dotted red curves indicate the background. In all panels, the long dashed orange curves are the $J=0$ signal. In the left panel, the dash-dotted violet curve is the $J=1$ signal, and the dashed green curve is the $J=2$ signal. In the right panel, the dash-dotted gray curve shows the combined $J=1$ and 2 signal.}
        \label{fig:2P_dataFit}
    \end{figure}
\end{center}

To determine the yield of reconstructed $\chi_{bJ}(2P)$ signal candidates, we define the shifted mass difference $\Delta M_{\chi}$ similarly to $\Delta M_{\pi\pi}$ (Eq.~\ref{eqn:dmpp}), as
\begin{equation}
    \label{eqn:dmc}
    \Delta M_{\chi} = M(\pi^{+}\pi^{-}\pi^{0}\ell^{+}\ell^{-}) - M(\ell^{+}\ell^{-}) + m\big(\Upsilon(1S)\big),
\end{equation}
where $M(\pi^{+}\pi^{-}\pi^{0}\ell^{+}\ell^{-})$ is the invariant mass of the $\chi_{bJ}$ final state. The distribution of signal events is narrowly peaked at the corresponding $\chi_{bJ}$ masses, with a corresponding resolution of $4.5-5.0~\text{MeV}$, depending on the transition. Signal yields are extracted from a simultaneous extended maximum-likelihood fit to the unbinned $\Delta M_{\chi}$ and $M_{\omega}$ distributions. Signal lineshapes are modeled with a double-sided Crystal Ball function (DSCB) \cite{Oreglia1980}, which consists of a Gaussian core that is augmented on either side by power-law tails. A single DSCB is used to parameterize the $M_\omega$ lineshape for the combined signals from the $J=1,2$ transitions, while for the distorted $M_{\omega}$ lineshape of the $J=0$ transition we use the product of a DSCB and a sigmoid. In the fit to data, the positions of the sigmoid midpoint and the $J=0$ DSCB centroid relative the $J=1,2$ DSCB centroid are fixed to the values extracted from simulation \cite{Stottler2022}. In the fit to data, the $\chi_{bJ}(2P)$ and $J=1,2$ $\omega$ mass are floated along with multiplicative resolution calibration factors $\rho$ and $\kappa$, which scale the Gaussian widths $\sigma$ of the DSCBs in $\Delta M_{\chi}$ and $M_{\omega}$, respectively. All remaining signal lineshape parameters $(\sigma, \alpha_i, n_i, b, \delta M)$ are fixed to values extracted from simulation. Backgrounds in $\Delta M_{\chi}$ and $M_{\omega}$ are modeled with cubic and quadratic polynomials, respectively. The projections of this fit are illustrated in Fig.~\ref{fig:2P_dataFit}, and the extracted signal yields are summarized in Table~\ref{tab:yields}.

\begin{table}[h!]
    \centering
    \caption{Selection efficiencies, obtained from large samples of simulation, as well as extracted signal yields in data and the associated significances, including systematic uncertainties, expressed in terms of standard deviations $(\sigma)$.}
    \begin{tabular}{lcrr}
        \hline\hline
        $\chi_{bJ}(2P)$~~ & ~~Efficiency (\%)~~ & ~~~~Signal Yield & ~~~~Significance \\
        \hline
        \noalign{\vskip 0.5mm}
        $J=0$ & $8.13\pm0.02$ & $32\pm11$ & $3.2\sigma$\\
        $J=1$ & $8.35\pm0.02$ & $304\asy{26}{24}$ & $16.5\sigma$\\
        $J=2$ & $8.36\pm0.02$ & $62\asy{17}{16}$ & $4.1\sigma$\\
        % $J=0$ & $8.13\pm0.02$ & $31.5\asy{11.2}{10.6}$ & $3.2\sigma$\\
        % $J=1$ & $8.35\pm0.02$ & $304\asy{26}{24}$ & $16.5\sigma$\\
        % $J=2$ & $8.36\pm0.02$ & $62.1\asy{16.9}{15.9}$ & $4.1\sigma$\\
        \noalign{\vskip 0.5mm}
        \hline\hline
    \end{tabular}
    \label{tab:yields}
\end{table}

\onecolumngrid
\twocolumngrid

As the $\chi_{bJ}(2P)$ are well established resonances, there is no look--elsewhere effect~\cite{Gross2010}; therefore, it is sufficient to calculate the local significance with masses fixed to their best-fit values while leaving the multiplicative resolution calibrations and signal yields free. We verify the applicability of Wilks' theorem \cite{Wilks1938} with a series of MC pseudoexperiments for each signal using the methodology described in Ref.~\cite{Chilikin2014}. The statistical significance, including systematic uncertainties, is calculated by convolving the profile likelihood~\cite{Cowan2011} with a Gaussian distribution of width equal to the systematic uncertainty of each signal hypothesis. The results are summarized in Table~\ref{tab:yields}. The $J=0$ peak has a significance of $3.2\sigma$; this is the first evidence for the near--threshold transition $\chi_{b0}(2P)\to\omega\Upsilon(1S)$.

\begin{figure}
    \centering
    \includegraphics[width=0.5\textwidth]{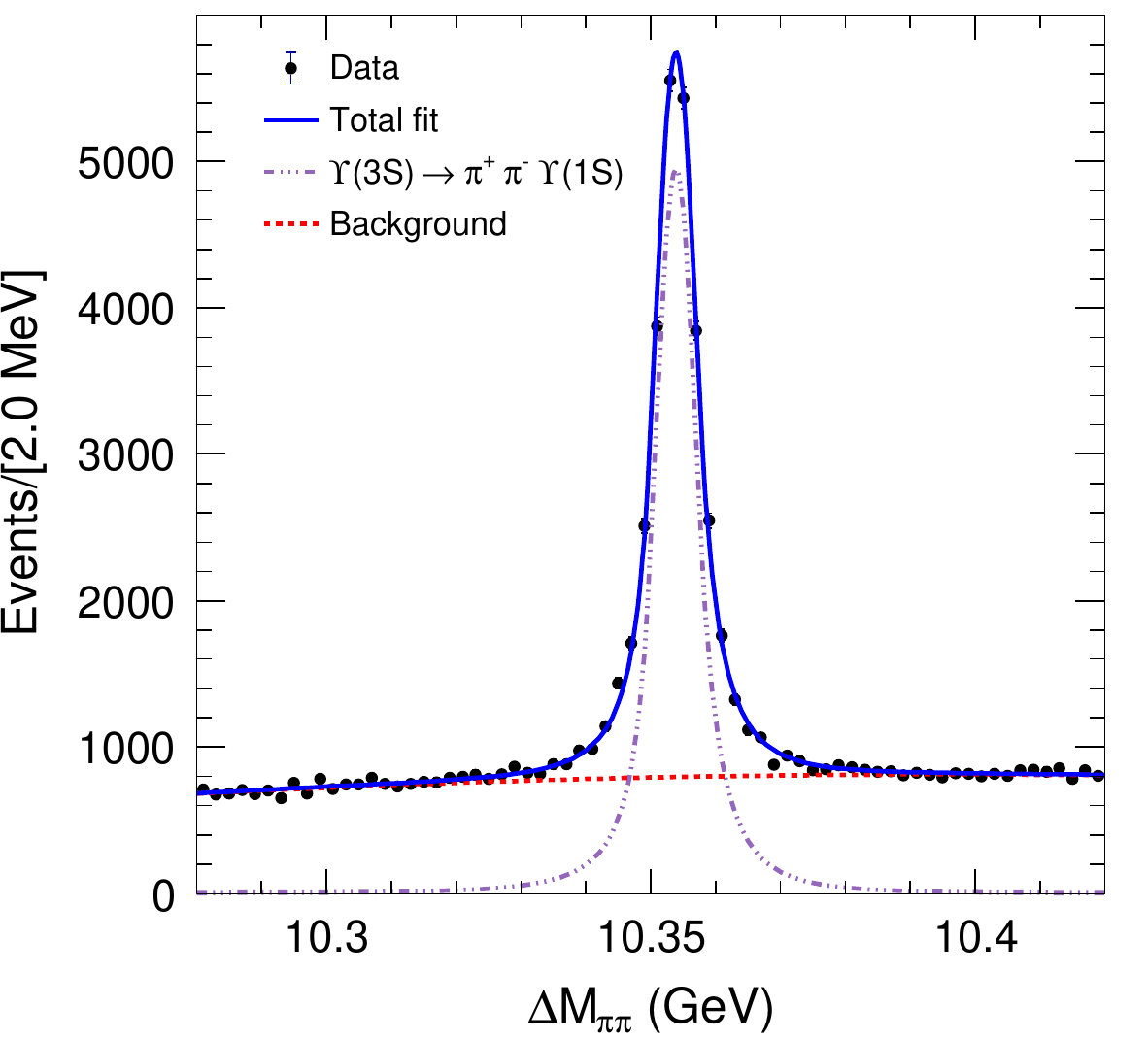}
    \caption{Fit to $\Delta M_{\pi\pi}$ for $\Upsilon(3S)\to\pi^+\pi^-\Upsilon(1S)$ events in the combined dataset. The total fit is shown by the solid blue curve, the background contribution by the dotted red curve, and the signal contribution by the dash-dotted violet curve is the signal shape.}
    \label{fig:dipi_dataFit}
\end{figure}

We attribute any $\chi_{bJ}(2P)$ in $\Upsilon(4S)$ data solely to radiative decays of ISR produced $\Upsilon(3S)$ mesons, as transitions of the form $\Upsilon(4S)\to X + \chi_{bJ}(2P)$ have not been seen~\cite{workman2022}. Branching fractions for the three $\omega$ transitions are calculated from the relevant signal yield $N_J$ and reconstruction efficiency $\epsilon_J$ as
\begin{equation}
  \label{eqn:2P_bfCalculation}
    \mathcal{B}\big(\chi_{bJ}(2P)\to\omega\Upsilon(1S)\big) =
    \frac{N_{J}}{\epsilon_{J} N_{\Upsilon(3S)} \prod \mathcal{B}_i},
\end{equation}
where $\prod\mathcal{B}_{i}$ is the product of the $\Upsilon(3S)\to\gamma\chi_{bJ}(2P)$, $\Upsilon(1S)\to\ell^{+}\ell^{-}$, $\omega\to\pi^{+}\pi^{-}\pi^{0}$, and $\pi^{0}\to\gamma\gamma$ branching fractions. The number of $\Upsilon(3S)$ events $N_{\Upsilon(3S)}$ is determined from a concurrent reconstruction of $\Upsilon(3S)\to\pi^{+}\pi^{-}\Upsilon(1S)[\ell^{+}\ell^{-}]$ events. Leptons and pions are reconstructed and combined to form di-pion and $\Upsilon(1S)$ candidates with selection criteria identical to those used to reconstruct $\chi_{bJ}(2P)\to\omega\Upsilon(1S)$ events, with the caveat that the constraint on the pion momentum is relaxed to $p_\text{c.m.}<0.7~\text{GeV}$. Additionally, the four charged tracks combine to form a shifted invariant mass $\Delta M_{\pi\pi}$ lying in the signal range of $[10.28,10.42]~\text{GeV}$. The resulting reconstruction efficiency $(\epsilon_{\pi\pi\Upsilon})$ is $(41.37\pm0.03)\%$. The $\pi\pi\Upsilon(1S)$ yield $N_{\pi\pi\Upsilon}$ is extracted from an extended maximum-likelihood fit to the unbinned data shown in Fig.~\ref{fig:dipi_dataFit}. The signal is parameterized with a DSCB and the background with a linear function. A signal yield of $24756^{+234}_{-233}\pm99$ events is obtained, where the first uncertainty is statistical and the second is systematic. The corresponding number of $\Upsilon(3S)$ events is calculated as:
\begin{equation}
    \label{eqn:N3S}
    \begin{split}
        N_{\Upsilon(3S)} &= \frac{N_{\pi\pi\Upsilon}}{\epsilon_{\pi\pi\Upsilon}\mathcal{B}\big(\Upsilon(3S)\to\pi^{+}\pi^{-}\Upsilon(1S)[\ell^{+}\ell^{-}]\big)},\\
        % &= \big(28.2\pm0.3\pm0.9\big)\times10^{6}.\\
    \end{split}
\end{equation}
yielding $\big(28.2\pm0.3\pm0.9\big)\times10^{6}$ events, where the statistical and systematic uncertainties are quoted, respectively.   The systematic uncertainty is the sum in quadrature of the following sources: tracking (1.3\%), lepton identification $(0.4\%)$, fit procedure $(0.4\%)$, input branching fractions $(3.1\%)$, and the efficiency $(0.1\%)$. The individual data samples contributing to this total are summarized in Table~\ref{tab:BE_datasets}.

    \begin{table}[htb]
        \centering
        \caption{Individual datasets which comprise the full sample of $\Upsilon(3S)$ decays studied in this analysis. The numbers of $\pi\pi\Upsilon$ events are obtained separately by fits to each sample, and to the combined sample.}
        \begin{tabular}{lcrr}
            \hline\hline
            \noalign{\vskip 0.5mm}  
            Dataset & \shortstack{$\mathcal{L}~(\text{fb}^{-1})$} & \multicolumn{1}{c}{$N_{\pi\pi\Upsilon}$} & \shortstack{Number of $\Upsilon(3S)$}\\
            \noalign{\vskip 0.5mm}
            \hline
            \noalign{\vskip 0.5mm}
            On-Res. $\Upsilon(3S)$ & 2.9 & $11852~\asy{113}{112}~$ & $(12.7\pm0.4)\times10^6$ \\
            Off-Res. $\Upsilon(4S)$ & 56.0~ & $1469~\asy{58}{57}~~$ & $(1.76\pm0.09)\times10^6$ \\
            On-Res. $\Upsilon(4S)$ & ~523.0 ~~~ & $11354~\asy{181}{180}~$ & $(13.6\pm0.5)\times10^6$\\
            \noalign{\vskip 0.5mm}
            \hline
            \noalign{\vskip 0.5mm}
            Combined & -- & $24756~\asy{234}{233}~$ & $(28.2\pm0.9)\times10^6$\\
            \noalign{\vskip 0.5mm}
            \hline\hline
        \end{tabular}
     
        \label{tab:BE_datasets}
    \end{table}
    
In the combination of Eqs.~\ref{eqn:2P_bfCalculation} and~\ref{eqn:N3S}, the $\Upsilon(1S)\to\ell^{+}\ell^{-}$ branching fraction cancels. Using the present world averages~\cite{workman2022}, we obtain
\begin{equation}
    \label{eqn:bfs_vals}
    \begin{split}
        \mathcal{B}\big(\chi_{b0}(2P)\to\omega\Upsilon(1S)\big) &= \big(0.55\pm0.19\pm0.07\big)\% \\
        \mathcal{B}\big(\chi_{b1}(2P)\to\omega\Upsilon(1S)\big) &= \big(2.39\asy{0.20}{0.19}\pm0.24\big)\% \\
        \mathcal{B}\big(\chi_{b2}(2P)\to\omega\Upsilon(1S)\big) &= \big(0.47\asy{0.13}{0.12}\pm0.06\big)\% \\
    \end{split}
\end{equation}
where the first uncertainty is statistical and the second is systematic. The $J=1,2$ measurements are consistent with the CLEO results~\cite{Pedlar2004} at less than $2\sigma$. It is worth noting that the $J=0$ and $J=2$ rates are consistent with one another, despite the broader natural width of the $J=0$ state and the limited phase space for the transition. There is experimental precedent for the enhancement of near-threshold $\omega$ transitions in the charmonium sector, where the the $\chi_{c1}(3872)$, previously dubbed the $X(3872)$ (an exotic meson candidate), decays to an $\omega J/\psi$ final state with a rate of $(5.0\pm1.9)\%$ \cite{workman2022}. Initially, the unexpectedly large $\omega J/\psi$ transition of the $\chi_{c1}(3872)$ was attributed to the exotic nature of the state. Given the new evidence for an analogous near-threshold transition from a conventional $b\bar{b}$ state presented in this Letter, it seems that the exotic nature of $\chi_{c1}(3872)$ may not provide the full explanation of the large branching fraction for $\chi_{c1}(3872)\to\omega J/\psi$. These decays evidently also occur due to a combination of the natural width of $\omega$, which enables them to proceed kinematically, and hadronic transition dynamics which are inadequately described by NRQCD.

Save for the $J=1$ measurement, the statistical uncertainty dominates. In obtaining the quoted systematic uncertainties, individual uncertainties have been assigned for the number of $\Upsilon(3S)~(1.0\%)$ and MC statistics $(0.2\%)$. An additional systematic uncertainty of 0.9\% is assigned to the efficiency of $\pi^{0}$ reconstruction by studying $\Upsilon(2S)\to\pi^0\pi^0\Upsilon(1S)$ events \cite{Stottler2022}. In the normalization procedure, correlated systematic uncertainties due to data-MC differences in track finding, lepton identification, and the $\Upsilon(1S)\to\ell^{+}\ell^{-}$ branching fraction cancel. The uncertainty on each signal yield due to the signal extraction procedure is estimated by repeating fits to data changing fit window, background parameterization, values of shape parameters fixed from signal MC events. To quantify the impact of uncertainties of fixed parameters on the extracted signal yields, the fit to data is repeated while varying the values of the fixed parameters, which are sampled from a multivariate Gaussian constructed from a block-diagonal global covariance matrix to account for correlation among the fixed parameters. The impact of the choice of background lineshape fit range are assessed by repeating the fit to data with higher-order polynomials and perturbed window boundaries. In both cases, the resulting distributions of extracted signal yields are fit with an asymmetric Gaussian and corresponding systematic uncertainties are calculated as the ratio of the widths and the fitted mean. \cite{Stottler2022}

The systematic uncertainty from each source is summed in quadrature to estimate the uncertainty assigned to signal extraction $\big(\asy{8.0\%}{8.3\%},~\asy{1.8\%}{2.0\%},~\text{and}~\asy{3.1\%}{4.6\%}$ for the $J=0$, 1, and 2 states, respectively$\big)$. The relative uncertainties on the branching fractions for $\Upsilon(3S)\to\gamma\chi_{bJ}(2P)$, $\Upsilon(3S)\to\pi^{+}\pi^{-}\Upsilon(1S)$, $\omega\to\pi^{+}\pi^{-}\pi^{0}$, and $\pi^{0}\to\gamma\gamma$ constitute the largest source of systematic uncertainty in the analysis $(10.4\%,~9.7\%,~\text{and}~12.4\%$ for the $J=0$, 1, and 2 states, respectively). These uncertainties are combined in quadrature to obtain the total systematic uncertainty on each measurement.

In the nonrelativistic limit of QCDME, the spin dependence of the decay amplitude $\Gamma\big(\chi_{bJ}(2P)\to\omega\Upsilon(1S)\big)$ factorizes. In the ratio $r_{2/1}$, defined in Eq.~\ref{eqn:rJ1}, the spin of the heavy quark decouples and the ratio of the $\omega\Upsilon(1S)$ decay amplitudes may be approximated as the ratio of the $S$--wave phase space factors $\Delta_{J} = M_{\chi_{bJ}(2P)} - M_{\Upsilon(1S)} - M_{\omega}$. Following the methodology presented in Ref.~\cite{MVoloshin2003}, we calculate this ratio using current world average values of bottomonium state masses and branching fractions~\cite{workman2022} as $r_{2/1}^\text{QCDME} = 0.77\pm0.16$, a $3.3\sigma$ shift from the previously reported value of $1.3\pm0.3$.

To measure $r_{J/1}$, we reparameterize the fit in terms of the signal yield ratios $\text{P}_{J/1} = N_{J}/N_{1}$, for $J=0,2$. Correcting the results for the relevant efficiencies as $r_{J/1}=\text{P}_{J/1}\big(\epsilon_{1}/\epsilon_{J}\big)$, we obtain $r_{0/1}=0.107\asy{0.038}{0.036}\pm0.009$ and $r_{2/1}=0.205\asy{0.062}{0.056}\asy{0.007}{0.010}$. In each ratio $r_{J/1}$, only the systematic uncertainties assigned for signal extraction and the selection efficiency (on each yield) contribute. This $r_{2/1}$ measurement differs from the expectation from QCDME~\cite{MVoloshin2003} by $3.3\sigma$. It should be noted that this test of QCDME is limited not by the statistics of our sample, but by the measured values of radiative $b\bar{b}$ branching fractions used in the calculation of $r_{2/1}^\text{QCDME}$.

In summary, we have used the combined $\Upsilon(3S)$ and $\Upsilon(4S)$ data samples collected by the Belle detector to obtain the first evidence for the near--threshold transition $\chi_{b0}(2P)\to\omega\Upsilon(1S)$ produced in radiative $\Upsilon(3S)$ decays with a branching fraction of $\big(0.55\pm0.19\pm0.07\big)\%$ at a significance of $3.2\sigma$. Moreover, we measure the hadronic transitions $\mathcal{B}\big(\chi_{b1}(2P)\to\omega\Upsilon(1S)\big) = \big(2.39\asy{0.20}{0.19}\pm0.24\big)\%$  and $\mathcal{B}\big(\chi_{b2}(2P)\to\omega\Upsilon(1S)\big) = \big(0.47\asy{0.13}{0.12}\pm0.06\big)\%$. This constitutes the first confirmation of the $J=1$ and $J=2$ branching fractions since their discovery~\cite{Pedlar2004}. The cascade branching fraction ratios $r_{J/1}$ are also measured. Comparison of the resulting measurement of $r_{2/1}$ with the theoretical expectations from QCDME---based on multipole expansions of $S$--wave phase space factors---reveals a $3.3\sigma$ tension.

%***** Acknowledgments *****
%----------- Long version, for most papers ----------- 
This work, based on data collected using the Belle detector, which was
operated until June 2010, was supported by 
the Ministry of Education, Culture, Sports, Science, and
Technology (MEXT) of Japan, the Japan Society for the 
Promotion of Science (JSPS), and the Tau-Lepton Physics 
Research Center of Nagoya University; 
the Australian Research Council including grants
DP210101900, % Urquijo
DP210102831, % Sevior
DE220100462, % Hsu
LE210100098, % Infrastructure
LE230100085; % Infrastructure
Austrian Federal Ministry of Education, Science and Research and
Austrian Science Fund (FWF) No.~P~31361-N36;
National Key R\&D Program of China under Contract No.~2022YFA1601903,
National Natural Science Foundation of China and research grants
No.~11575017,
No.~11761141009, 
No.~11705209, 
No.~11975076, 
No.~12135005, 
No.~12150004, 
No.~12161141008, 
No.~12175041, 
and
No.~12475093,
and Shandong Provincial Natural Science Foundation Project ZR2022JQ02;
the Czech Science Foundation Grant No. 22-18469S;
Horizon 2020 ERC Advanced Grant No.~884719 and ERC Starting Grant No.~947006 ``InterLeptons'' (European Union);
the Carl Zeiss Foundation, the Deutsche Forschungsgemeinschaft, the
Excellence Cluster Universe, and the VolkswagenStiftung;
the Department of Atomic Energy (Project Identification No. RTI 4002), the Department of Science and Technology of India,
and the UPES (India) SEED finding programs Nos. UPES/R\&D-SEED-INFRA/17052023/01 and UPES/R\&D-SOE/20062022/06; 
the Istituto Nazionale di Fisica Nucleare of Italy; 
National Research Foundation (NRF) of Korea Grant
Nos.~2021R1\-F1A\-1064008,
2022R1\-A2C\-1003993,
2022R1\-A2C\-1092335,
RS\-2016\-NR017151,
RS\-2018\-NR031074,
RS\-2021\-NR060129,
RS\-2023\-00208693,
RS\-2025\-02219521;
Radiation Science Research Institute, Foreign Large-size Research Facility Application Supporting project, the Global Science Experimental Data Hub Center, the Korea Institute of Science and Technology Information (K25L2M2C3) and KREONET/GLORIAD;
the Polish Ministry of Science and Higher Education and 
the National Science Center;
the Ministry of Science and Higher Education of the Russian Federation
and the HSE University Basic Research Program, Moscow; % from 15.04.2021
University of Tabuk research grants
S-1440-0321, S-0256-1438, and S-0280-1439 (Saudi Arabia);
the Slovenian Research Agency Grant Nos. J1-9124 and P1-0135;
Ikerbasque, Basque Foundation for Science, and the State Agency for Research
of the Spanish Ministry of Science and Innovation through Grant No. PID2022-136510NB-C33 (Spain);
the Swiss National Science Foundation; 
the Ministry of Education and the National Science and Technology Council of Taiwan;
and the United States Department of Energy and the National Science Foundation.
We thank the KEKB group for the excellent operation of the
accelerator; the KEK cryogenics group for the efficient
operation of the solenoid; and the KEK computer group and the Pacific Northwest National
Laboratory (PNNL) Environmental Molecular Sciences Laboratory (EMSL)
computing group for strong computing support; and the National
Institute of Informatics, and Science Information NETwork 6 (SINET6) for
valuable network support.
This material is based in part on work supported by the U.S. Department of Energy,
Office of Science, Office of Workforce Development for Teachers and Scientists,
Office of Science Graduate Student Research (SCGSR) program. The SCGSR program is
administered by the Oak Ridge Institute for Science and Education (ORISE) for the DOE.
ORISE is managed by ORAU under contract number DE-SC0014664.

These acknowledgements are not to be interpreted as an endorsement of any
statement made by any of our institutes, funding agencies, governments, or
their representatives.

\bibliography{bibliography}

\end{document}

%% file: pub649-orcid.tex
%%% Paper:    chi_bJ(2P) to omega Y(1S)
%%% Journal:  Physical Review Letters
%%% Contacts: Z. Stottler, T. Pedlar
%%% Non-responding authors or those who said NO are commented out.
%%% ====================================================================
%%% Click the RELOAD button on your web browser to see the updated file.
%%% ====================================================================
%%% Use \input{pub649-orcid} to insert this material into your latex file.
%%%%% Force institutions to appear in alphabetical order when typeset.
\noaffiliation
  \author{Z.~S.~Stottler\,\orcidlink{0000-0002-1898-5333}} % 2267
  \author{T.~K.~Pedlar\,\orcidlink{0000-0001-9839-7373}} % 2421
  \author{B.~G.~Fulsom\,\orcidlink{0000-0002-5862-9739}} % 2563
  \author{I.~Adachi\,\orcidlink{0000-0003-2287-0173}} % 2590
  \author{K.~Adamczyk\,\orcidlink{0000-0001-6208-0876}} % 2239
% \author{J.~K.~Ahn\,\orcidlink{0000-0002-5795-2243}} % 7423
  \author{H.~Aihara\,\orcidlink{0000-0002-1907-5964}} % 2223
  \author{S.~Al~Said\,\orcidlink{0000-0002-4895-3869}} % 6823
  \author{D.~M.~Asner\,\orcidlink{0000-0002-1586-5790}} % 4684
  \author{H.~Atmacan\,\orcidlink{0000-0003-2435-501X}} % 2538
% \author{V.~Aulchenko\,\orcidlink{0000-0002-5394-4406}} % 8183
  \author{T.~Aushev\,\orcidlink{0000-0002-6347-7055}} % 3747
% ...
% 2387
% \author{V.~Zhulanov\,\orcidlink{0000-0002-0306-9199}} % 4983
\collaboration{The Belle Collaboration}

%% file: pub649.bbl
\begin{thebibliography}{37}
\bibitem [{uni()}]{unitsFootnote}%
  \BibitemOpen
  \href@noop {} {}\bibinfo {note} {We use units in which the speed of light is $c=1$.}\BibitemShut {Stop}%
\end{thebibliography}
